\documentclass[aps,prb,twocolumn,showpacs,floats,graphicx,letterpaper,superscriptaddress]{revtex4-1}
\pdfoutput=1

\usepackage{graphicx}
\begin{document}


\title{Nonequilibrium Quasiparticle Relaxation Dynamics in Single Crystals of Hole and Electron doped BaFe$_2$As$_2$}

\author{Darius H Torchinsky}
\author{James W. McIver}
\author{David Hsieh}
\affiliation{Department of Physics, Massachusetts Institute of
Technology, Cambridge, Massachusetts, 02139, USA}

\author{G.F. Chen}
\author{J.L. Luo}
\author{N. L. Wang}
\affiliation{Beijing National Laboratory for Condensed Matter
Physics, Institute of Physics, Chinese Academy of Sciences,
Beijing 100190, China}

\author{Nuh Gedik}%
\email{gedik@mit.edu} \affiliation{Department of Physics,
Massachusetts Institute of Technology, Cambridge, Massachusetts,
02139, USA}

\date{\today}

\begin{abstract}
We report on the nonequilibrium quasiparticle dynamics in
BaFe$_2$As$_2$ on both the hole doped (${\rm
Ba_{1-x}K_{x}Fe_2As_2}$) and electron doped (${\rm
BaFe_{2-y}Co_{y}As_2}$) sides of the phase diagram using ultrafast
pump-probe spectroscopy. Below $T_c$, measurements conducted at
low photoinjected quasiparticle densities in the optimally and
overdoped ${\rm Ba_{1-x}K_{x}Fe_2As_2}$ samples reveal two
distinct relaxation processes: a fast component whose decay rate
increases linearly with excitation density and a slow component
with an excitation density independent decay rate. We argue that
these two processes reflect the recombination of quasiparticles in
the two hole bands through intraband and interband processes. We
also find that the thermal recombination rate of quasiparticles
increases quadratically with temperature in these samples. The
temperature and excitation density dependence of the decays
indicates fully gapped hole bands and nodal or very anisotropic
electron bands. At higher excitation densities and lower hole
dopings, the dependence of the dynamics on quasiparticle density
disappears as the data are more readily understood in terms of a
model which accounts for the quasiequilibrium temperature attained
by the sample. In the ${\rm BaFe_{2-y}Co_{y}As_2}$ samples,
dependence of the recombination rate on quasiparticle density at
low dopings (i.e., $y=0.12$) is suppressed upon submergence of the
inner hole band and quasiparticle relaxation occurs in a slow,
density independent manner.
\end{abstract}

\maketitle

\section{\label{sec:intro}Introduction}

The discovery of the iron-pnictide
superconductors\cite{Kamihara2008,Rotter2008} has opened a new
chapter in the study of high-temperature superconductivity.
Although they display a lower degree of electronic correlation
than the cuprates\cite{Qazilbash2009}, there are a number of
similarities between the two material classes. Perhaps the most
striking of these is the close correspondence between their phase
diagrams. In both classes, superconductivity is achieved by the
doping of either electrons or holes into a ``parent'' material
which displays magnetic ordering at lower temperatures, i.e., spin
density wave (SDW) ordering in the
pnictides\cite{DelaCruz2008,Huang2008} or antiferromagnetic
ordering in the cuprates. As this static magnetic ordering is
suppressed with doping, superconductivity emerges, increasing in
transition temperature $T_c$ to a maximum at optimal doping before
receding with the addition of further carriers.

With these shared characteristics come shared mysteries, including
the natures of the pseudogap state, of the coexistence between
magnetic and superconducting order (i.e., homogeneous vs.
heterogeneous), and, arguably the most fundamental, the binding
boson itself. The presence of a resonance in neutron scattering
spectra below $T_c$ in the pnictides\cite{christianson,lumsden}
and the suppression of magnetic ordering upon the addition of
carriers suggests that antiferromagnetic spin fluctuations mediate
pairing, implying that the superconducting order parameter
$\Delta$ switches sign in the Brillouin zone. Therefore,
definitive determination of the gap structure and symmetry would
serve as a significant step towards understanding the mechanism of
superconductivity. In the cuprates, experiments clearly show
$d$-wave pairing\cite{tsuie2}. The picture in the pnictides is a
lot murkier, in part due to their multiband character and the fact
that interband interactions likely play a role in
superconductivity\cite{mazin,wang2009,chubukov}. This complexity
may be at the origin of the seemingly conflicting sets of
experimental evidence concerning the symmetry of $\Delta$. Angle
Resolved Photoemission (ARPES) studies have consistently revealed
a fully gapped Fermi surface\cite{ding,ding2,wray}, while
NMR\cite{fukazawa} and magnetic penetration depth\cite{martin}
measurements suggest the presence of nodes. Confronted with these
data, theorists have proposed a diverse collection of order
parameters, including an $s_{\pm}$ symmetry with
nodes\cite{graser}, $d_{xy}$ symmetry \cite{yanagi}, and the
prevailing nodeless $s_{\pm}$ order parameter\cite{mazin} with
interband impurity scattering\cite{chubukov}.

In this study, we use optical pump-probe spectroscopy to
investigate the gap symmetry and interband interactions in the
Ba-122 system of the iron pnictides. Pump-probe spectroscopy is a
powerful technique in which absorption of an ultrashort ``pump''
pulse results in the injection of a transient population density
$n$ of nonequilibrium quasiparticles\cite{parker1972}. A second
``probe'' pulse then records the return of the system to
equilibrium through measurement of the time-resolved change in
reflectivity ($\Delta R(t)/R$), assumed linearly proportional to
$n$. These measurements have been successfully employed in the
cuprates to examine, e.g., gap symmetry\cite{gedik,gedik2} and the
strength of electron-boson coupling \cite{demsar,gadermaier}. In
the pnictides, pump-probe spectroscopy has probed the existence of
a pseudogap state\cite{mertelj}, the competition between SDW and
SC ordering\cite{chia}, coherent lattice vibrations to rule out
their role in superconductivity\cite{mansart}, and the role of
interband interactions and gap symmetry in nonequilibrium
quasiparticle relaxation\cite{torchinsky}. As prior ultrafast
experiments have focused on one side of the phase diagram at a
time, there is a need for a systematic study of the pnictides as a
function of doping across the phase diagram.

Building upon our prior work\cite{torchinsky}, we study both the
hole-doped Ba$_{1-x}$K$_x$Fe$_2$As$_2$ and the electron-doped
BaFe$_{2-y}$Co$_y$As$_2$ series of the 122 pnictides. In the
optimal to overdoped hole doped samples ($x=0.4,0.5,0.6$), the
signal ($\Delta R(t)/R$) is composed of two distinct features. The
first is a fast component whose decay rate depends on excitation
fluence $\Phi$ which, aided by ARPES\cite{ding,richard,wray}
measurements and LDA calculations\cite{ma}, we argue arises from
the inner hole bands. The second is a slow, fluence independent
decay we attribute to the outer hole band. An analysis of the fast
component based on the Rothwarf-Taylor\cite{roth-tay,gedik}
coupled differential equations reveals a $T^2$ dependence of the
thermal population of quasiparticles on temperature, suggesting
the presence of nodes on the Fermi Surface. At the same time, the
number of photoinduced quasiparticles that contribute to the
overall signal increases linearly with laser fluence, reflecting
the fully gapped character of these bands. We thus argue that
these observations are consistent with fully gapped hole pockets
at zone center and nodal or highly anisotropic electron pockets at
the zone boundary.

In the normal state, we observe oscillations in the reflectivity
transients due to coherent acoustic phonons. These thermally
driven acoustic waves are suppressed below $T_c$ due to a
significant decrease in the thermal expansion coefficient below
the transition\cite{budko,hardy,daluz} -- except at the highest
fluences, suggesting a dichotomy between the low and high fluence
regimes.

These arguments provide a framework for understanding the rest of
the available phase diagram. At lower hole doping levels (i.e.,
$x=0.2$, $x=0.3$) and higher excitation fluences, the data
represent the normal-state behavior seen at higher dopings. In the
low fluence regime, however, these samples do not exhibit the
density-dependent recombination seen for the optimally and
overdoped samples, precluding an analysis to determine Fermi
surface topology. Rather, these data are qualitatively explained
by a simple model that accounts for the spatially integrated
nature of the pump-probe measurement.

In the Co-doped samples, low-fluence measurements below $T_c$ in
an underdoped ($y=0.12$) sample indicate quasiparticle
recombination that is weakly dependent on incident fluence. This
intensity dependent recombination disappears as further electron
doping submerges the innermost hole pocket below the Fermi level;
for large values of Co-doping, the only observed relaxation is the
slow decay we previously assigned to the outer hole band.

\section{\label{sec:setup}Experimental Setup}

We employed a Ti:sapphire oscillator system which produces 60~fs
pulses at central wavelength 795~nm (photon energy 1.5~eV) at an
80~MHz repetition rate. After passing through a prism compressor
designed to account for the dispersion of the setup, the
repetition rate of the laser was reduced to 1.6~MHz with an
electro-optic pulse-picker in order to eliminate the effects of
cumulative heating of the sample. The beam was then split into two
paths with a 9:1 ratio between the pump and probe. The
horizontally polarized pump beam was chopped at 102~kHz by a
photoelastic modulator/polarizer pair for lock-in detection and
then passed through neutral density filters in order to tune the
incident pump fluence from ${\rm 44~\mu J/cm^2}$ to ${\rm
44~nJ/cm^2}$. After attenuation, the pump beam was passed through
a fast scanning delay stage which swept the pump delay by
$\sim$40~ps at a rate of 30~Hz before being focused to a ${\rm
60~\mu m}$ FWHM spot-size on the sample.

After being split from the pump, the probe beam passed through a
computer-controlled motorized delay stage and was rotated to
vertical polarization before being focused to a ${\rm 60~\mu m}$
FWHM spot-size on the sample in the same location as the pump. The
reflected probe beam was separated from the pump by spatial and
polarization filtering, then focused onto a photodiode. The
photodiode output was low-pass filtered and measured by a lockin
amplifier synched with the photoelastic modulator. We used a fast
time constant (${\rm 30~\mu s}$) so that the lockin output could
be input to an oscilloscope triggered to the fast scanning stage.
We averaged several fast-scan stage sweeps to record each data
trace.

All samples were grown using the self-flux method in proportions
given by their chemical formula. A variety of sample
characterization methods were employed with the most rigorous
testing reserved for the optimally hole-doped ($x=0.4$) sample as
follows: the induction-coupled plasma (ICP) technique was used to
determine the Ba/K ratio, the uncertainty of which was determined
to be less than 2\%\cite{chen}. Energy-dispersive X-ray (EDX)
spectroscopy was used to determine spatial variation in sample
doping homogeneity. A plot of the resulting EDX analysis pattern
may be found in Figure~1b of Ref.~\onlinecite{chen}. In a recent
EDX analysis on a different batch of samples, the maximum spatial
variation was found to range from $x=0.39$ to $x=0.44$.
Resistivity and magnetic susceptibility measurements have been
independently performed and presented\cite{wray}; the sharpness of
the transition was used as a gauge of sample purity. These results
are shown in Figs.~1a and 1b of Ref.~\onlinecite{wray}.

A number of other studies on samples from the same batch of
optimally-hole doped compounds have been performed, including
ARPES\cite{wray,ding,richard}, Infrared spectroscopy\cite{li}, and
thermal conductivity\cite{Checkelsky}; further information
regarding sample purity and homogeneity may be found therein. We
performed SQUID magnetometry on all the samples used in this study
and measured $\Delta T_c\sim 1$~K at each sample's transition
temperature.

\section{\label{sec:resdisc}Results and Discussion}

\begin{figure}
\includegraphics[scale=1]{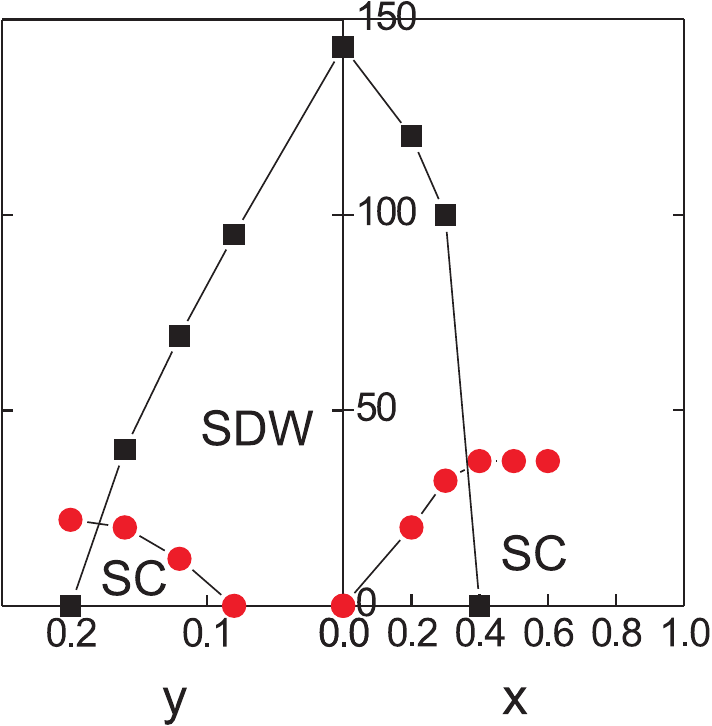}
\caption{\label{fig:phase} Schematic phase diagram of the Ba-122
pnictides for the hole-doped Ba$_{1-x}$K$_x$Fe$_2$As$_2$ and
electron-doped BaFe$_{2-y}$Co$_y$As$_2$ samples used in this
study. The $T_c$ values are denoted by black squares and the
$T_{SDW}$ values are given by red circles. All $T_c$ values were
measured by SQUID magnetometry. The values for $T_{SDW}$ on the
electron-doped side of the phase diagram were determined from the
same batch of samples\cite{nlwang}, while the values for $T_{SDW}$
on the hole-doped side of the phase diagram are from the
literature\cite{chen2}.}
\end{figure}

A schematic representation of the Ba-122 phase diagram is
presented in Fig.~\ref{fig:phase}. The data points indicate the
$T_c$ and $T_{SDW}$ values for the samples used in this study.
Specifically, we examined hole-doped Ba$_{1-x}$K$_x$Fe$_2$As$_2$
with $x=0.2-0.6$ in increments of 0.1, spanning the regime from
underdoping to overdoping. On the electron-doped side of the phase
diagram, we studied BaFe$_{2-y}$Co$_y$As$_2$ samples at doping
levels of $y=0.08-0.20$ in increments of 0.04, which spans
underdoped to optimally doped samples. The parent compound was
also measured.

We start by describing our measurements in
Ba$_{1-x}$K$_x$Fe$_2$As$_2$ in the optimally ($x=0.4$) to
overdoped ($x=0.6$) regime at the low excitation densities.
Shifting our attention to the high fluence regime provides a
framework for measurements in the underdoped $x=0.2,0.3$ samples,
which we follow with a discussion of measurements in the parent
compound and BaFe$_{2-x}$Co$_x$As$_2$ at all dopings.

\subsection{\label{sec:bakfeaso}Ba$_{1-x}$K$_x$Fe$_2$As$_2$: The Low Fluence Regime}

\begin{figure}
\includegraphics[scale=1]{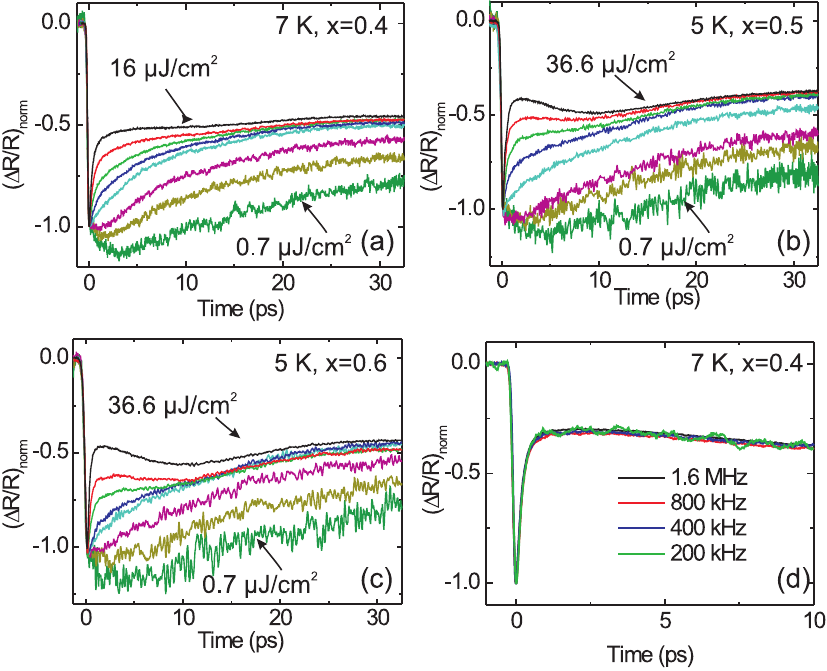}
\caption{\label{fig:rawdata}Fast, intensity-dependent relaxation
of the normalized reflectivity transients $(\Delta R/R)_{norm}$ in
the superconducting state. (a) The decay rate of $(\Delta
R(t)/R)_{norm}$ in the optimally doped compound ($x=0.4$)
systematically decreases with decreasing intensity at 7~K (from
bottom to top: $\Phi=$ 0.7~${\rm \mu J/cm^2}$ (light green), 2.2,
4.4, 7.0, 8.9, 12.8 to 16.1~${\rm \mu J/cm^2}$ (black)). Raw
traces from the overdoped hole-doped samples
Ba$_{1-x}$K$_x$Fe$_2$As$_2$ for (b) $x=0.5$ and (c) $x=0.6$ at $T
= 5$~K are qualitatively identical to those for the optimally
doped sample in panel (a). In both of these plots, the absorbed
fluences are, from bottom to top, $\Phi=$ 0.7~${\rm \mu J/cm^2}$
(blue), 1.4, 2.2, 4.4, 7.0, 12.7, 20.1, and 36.6~${\rm \mu
J/cm^2}$ (black). (d) $(\Delta R/R)_{norm}$ measured at four
different repetition rates are identical, verifying the absence of
cumulative heating ($\Phi = 37~{\rm\mu J/cm^2}$, $T$=7~K).}
\end{figure}

Characteristic short-time traces of the normalized change in
reflectivity $\Delta R/R$ as a function of time at various
fluences $\Phi$ are shown in Fig.~\ref{fig:rawdata}a for the
$x=0.4$ sample at $T$=7~K. There is an initial decrease of the
reflectivity at the arrival of the pump beam at time $t=0$.
Recovery to equilibrium depends strongly on the incident fluence
with higher fluences exhibiting a faster initial relaxation rate
than lower ones. Qualitatively identical intensity dependence is
observed in the $x=0.5,0.6$ samples, as shown in
Figs.~\ref{fig:rawdata}b and \ref{fig:rawdata}c. In all cases,
higher fluences are marked by faster initial relaxation of the
signal than for lower fluences. We note the presence of the
beginning of a heavily damped oscillation in the highest fluences
of this Figure (not shown for the $x=0.4$ sample, although it is
was observed) due to Stimulated Brillouin Scattering (SBS) which
disappears with decreasing fluence. This characteristic of the
data will be described in further detail below within the context
of the high-fluence regime measurements.

In order to rule out steady state heating as the source of this
intensity dependence, we used a pulse picker to vary the
repetition rate. Figure~\ref{fig:rawdata}d shows 7 K transients
taken with an absorbed pump fluence $\Phi = 37~\mu$J/cm$^2$ at
repetition rates ranging from 200~kHz to 1.6~MHz. We observe no
discernible change in the recovery dynamics. Measurements
performed above $T_c$ exhibited an absence of the intensity
dependence of Figs.~\ref{fig:rawdata}a-c, verifying its origin in
superconductivity\cite{torchinsky}.

\begin{figure}
\includegraphics[scale=1]{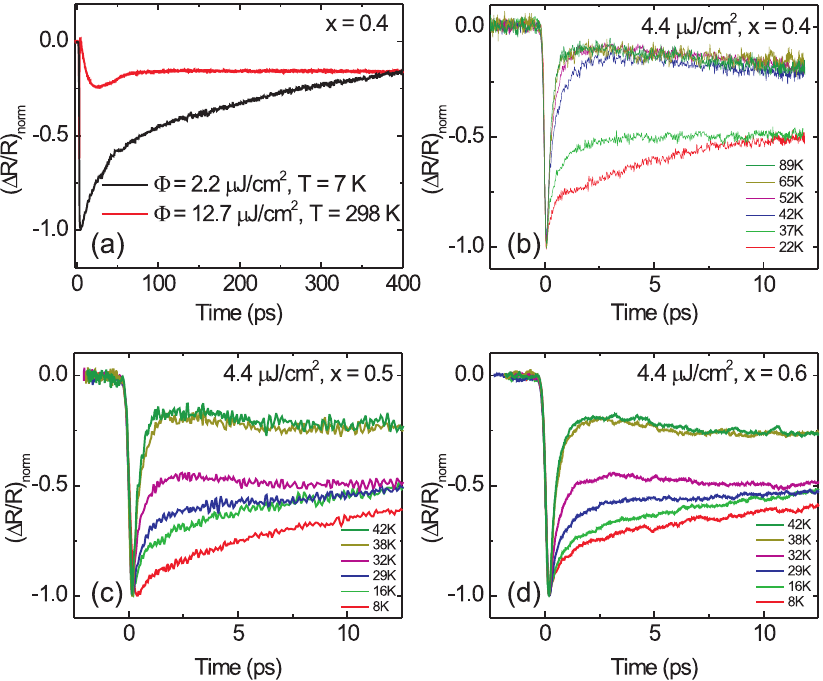}
\caption[Slow component]{\label{fig:slow} Slow relaxation of
$(\Delta R/R)_{norm}$ in the superconducting state. (a) $(\Delta
R/R)_{norm}$ at 7~K and at room temperature. The slow, excitation
density independent portion (black) switches off to become a
nearly flat response in the normal state, which clearly shows the
first oscillation due to stimulated Brillouin scattering. (b-d) A
more complete view of the temperature dependence of $(\Delta
R/R)_{norm}$ near $T_c$ obtained at $\Phi=4.4~J/cm^2$ in the
$x=0.4-0.6$ samples. We note a sharp decrease in the offset across
the transition (i.e. 37~K to 42~K) due to switching off of the
long-time component with loss of superconductivity. Above $T_c$
there is an upturn of the signal evident at short times due to
stimulated Brillouin scattering \cite{thomsen}.}
\end{figure}

The reflectivity transients of Figs.~\ref{fig:rawdata}a-c tend to
an offset at the end of the 30~ps measurement time window. This
offset is the beginning of a slow, intensity-independent decay
with a characteristic decay time of $\sim$500~ps\cite{torchinsky},
represented for the $x=0.4$ sample at 7~K by the black trace in
Fig.~\ref{fig:slow}a. As with the fast, intensity dependent
component, the slowly decreasing offset is observed to abruptly
shut off for all samples at their respective transition
temperatures, leading to a much longer decay of lower magnitude,
as seen in the red trace of Fig.~\ref{fig:slow}a. This switching
off of the long-time component was also a universal characteristic
of the $x=0.4-0.6$ samples, which is demonstrated by
Figs.~\ref{fig:slow}b-d.

\begin{figure}
\includegraphics[scale=1]{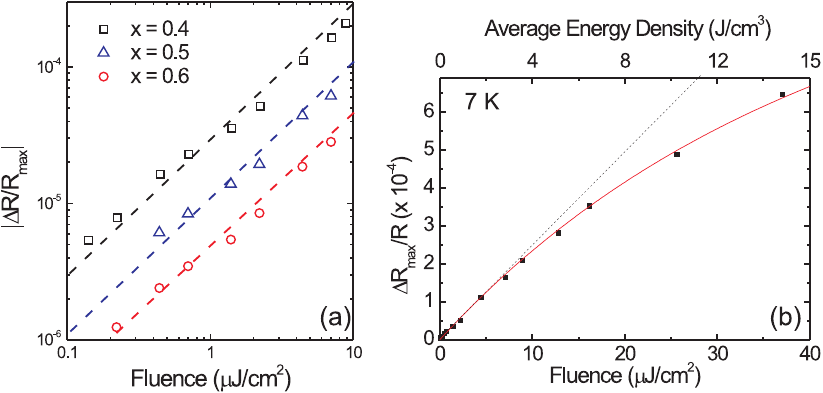}
\caption{\label{fig:saturation} (a) $|(\Delta R /R)_{max}|$
plotted as a function of $\Phi$ at $T=$7~K for the $x=0.4-0.6$
samples, each offset for clarity. Dashed lines show a slope of
one. Linear behavior in the low fluence regime is an indication
that the optically accessed excitations are fully gapped, as
described in the text. (b) The absolute value of the maximum
change in reflectivity versus both $\Phi$ and the average energy
absorbed within a penetration depth for the $x=0.4$ sample. The
red curve is a fit to a simple saturation model yielding
$\Phi_{sat}$ = 25~${\rm \mu J/cm^2}$ at an initial temperature of
7~K.}
\end{figure}

Further insight into the data is obtained by considering the
initial amplitude of the reflectivity transients.
Fig.~\ref{fig:saturation}a presents $(\Delta R(0)/R)$ as a
function of the absorbed pump fluence in the low-fluence regime at
$T=7$~K in the $x=0.4-0.6$ samples. Lines of slope 1 are shown
with the data, indicating that at these fluences, $\Delta R(0)/R
\propto n_0$ is directly proportional to $\Phi$. This linear
proportionality signifies that these experiments represent the
low-fluence regime. Significantly, a linear dependence of $n_0$ on
$\Phi$ is an indication that the photoinduced excitations are
fully gapped; the number of excitations is simply proportional to
the laser energy absorbed \cite{gedik}. In the presence of a line
node in the gap of the probed excitations, a linear dependence of
the density of states $g(E)$ on energy $E$ produces an estimate
for the total energy absorbed by the quasiparticles as $E_T
\sim\int g(E) EdE\sim E^3$. In such a case, the total number of
excited quasiparticles $n_{ph}$ is given as $n_{ph}\sim\int g(E)dE
\sim E^2$. Thus, the number of quasiparticles in terms of the
energy deposited is given as $n_{ph}\sim E_T^{2/3}$. Thus, a line
node in the vicinity of the photoexcited quasiparticles should
yield a sublinear dependence of the initial reflectivity
transients $\Delta R(0)/R \propto \Phi^{2/3}$, which is not
observed.

With increasing $\Phi$, $\Delta R(0)/R$ displays saturation
behavior at high excitation densities, as seen for the $x=0.4$
sample 7~K data in Fig.~\ref{fig:saturation}b. We fit these data
to a simple saturation model which accounts for the exponential
penetration of the light into the sample, given by
\begin{equation}
\Delta R(0)/\Delta R_{sat}=\lambda ^{-1}\int e^{-z/\lambda}
(1-e^{-\Phi(z)/\Phi_{sat}})dz
\end{equation}
where $\Phi(z) =\Phi(0)e^{-z/\lambda}$ is the laser fluence at a
depth of $z$ beneath the sample surface, $\Delta R_{sat}$ is the
change in reflectivity at saturation and $\lambda$ is the optical
penetration depth (26~nm in optimally doped ${\rm
Ba_{1-x}K_{x}Fe_2As_2}$ \cite{li}). The fit yields a saturation
fluence of $\Phi_{sat}=25~{\rm \mu J/cm^2}$ and $\Delta R_{sat}/R$
= $1.34~\times 10^{-3}$. Below, we restrict our analysis to $\Phi
< 2.5~{\rm \mu J/cm^2}$ which is clearly in the linear range of
Fig.~\ref{fig:saturation}b for all three dopings.

In the low-excitation regime, the dynamics of the photoinduced
quasiparticles may be understood within the framework of the
Rothwarf-Taylor model. This phenomenological model was originally
developed to describe nonequilibrium quasiparticle recombination
in tunneling experiments\cite{roth-tay}. Here, the coupled
recovery of a nonequilibrium concentration of quasiparticles $n$
and their binding boson $N$ may be described by the rate equations
\begin{eqnarray}\label{eq:rothtay}
\dot{n}&=&I_{qp}+2\gamma_{pc}N-B n^2\\
\dot{N}&=&I_{ph}+B n^2/2 -\gamma_{pc}N-(N-N_{eq})\gamma_{esc},
\end{eqnarray}
where, $I_{qp}$ and $I_{ph}$ are the external quasiparticle and
boson generation rates, respectively, $\gamma_{pc}$ is the pair
creation rate via annihilation of gap energy bosons, $N_{eq}$ is
the equilibrium boson number density, $\gamma_{esc}$ is the boson
escape rate, and $B$ is the bi-molecular recombination constant.
The rates $B n$, $\gamma_{pc}$ and $\gamma_{esc}$ describe the
three different relaxation pathways for the energy deposited into
the system by the ultrashort pulse.

\begin{figure}
\includegraphics[scale=1]{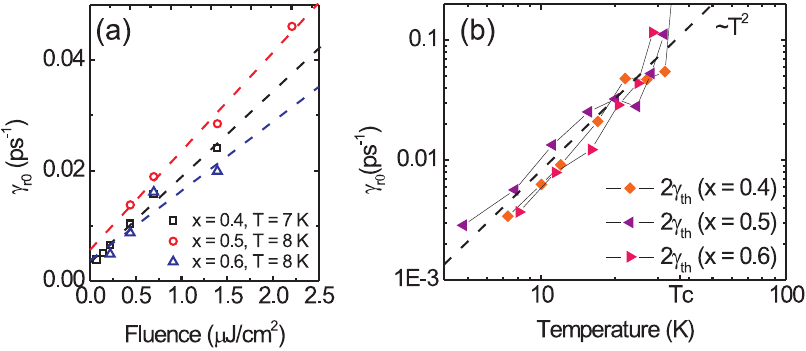}
\caption{\label{fig:Medusa}Decay-rate analysis of data with a view
towards gap symmetry. (a) The initial decay rate ($\gamma_{r0}$)
as a function of pump fluence at 7, 8~K for $x = 0.4 - 0.6$. The
linear dependence of recombination rate on initial population is a
hallmark of bimolecular recombination. Dashed lines show linear
fits to data. The thermal decay rate at each temperature was
obtain by extrapolating the fits to zero fluence. There is little
significant difference between the various dopings. (b) The
thermal relaxation rate $\gamma_{th}$ plotted as a function of
temperature for $x=0.4-0.6$. The dashed line shows $\sim T^2$
dependence as a guide along for all doping levels, suggesting a
negligible change in Fermi surface topology between differently
doped samples.}
\end{figure}

Depending on the relative magnitudes of these three rates, the
solutions of these equations display different characteristics. In
the simplest case, if $\gamma_{pc} \ll B n$ or $\gamma_{esc} \gg
\gamma_{pc}$ then the pair creation term in the first equation can
be ignored and the two equations decouple. In this limit,
quasiparticles display simple bimolecular kinetics governed by the
recombination coefficient $B$, as is observed, e.g., in
Figs.~\ref{fig:rawdata}a-c. In the limit where $\gamma_{esc} \ll
\gamma_{pc},B n$, quasiparticles and bosons come to a
quasiequilibrium and the combined population decays with a slow
rate proportional to $\gamma_{esc}$. This regime, known as the
phonon bottleneck, is consistent with the slow component of the
signal, shown in Fig.~\ref{fig:slow}a.

In the decoupled regime, the RT equations permit determination of
the thermal decay rate of quasiparticles ($\gamma_{th}$) and the
recombination coefficient $B$\cite{gedik}. This may be seen by
explicitly drawing a distinction between photoinduced
quasiparticles ($n_{ph}$) and thermally present ones ($n_{th}$)
and rewriting $n=n_{ph}+n_{th}$. Eq.~\ref{eq:rothtay} then becomes
\begin{equation}\label{eq:rt-lin}
dn/dt=-B n_{ph}^2-2 B n_{ph} n_{th},
\end{equation}
where the second term $Bn_{ph} n_{th}$ represents the
recombination of photoinduced quasiparticles with their thermally
populated counterparts. We consider the initial recombination rate
$\gamma_{r0}$, defined as
\begin{eqnarray}
\gamma_{r0}&=&-(1/n_{ph})(dn_{ph}/dt)|_{t\rightarrow 0}\\
\label{eq:therm}&=&B n_{ph} +2B n_{th}=\gamma_{ph}+2\gamma_{th}
\end{eqnarray}
where $\gamma_{ph,th} = B n_{ph,th}$ are the photoinduced and
thermal decay rates.

Experimentally, we deduce $\gamma_{r0}$ from the initial slope of
the reflectivity transients following the peak. As suggested by
Eq.~\ref{eq:therm}, we obtain $\gamma_{th}$ by extrapolating the
linear fits to zero excitation density (i.e., $n_{ph} \propto \Phi
\rightarrow 0$) for each temperature, as shown in
Fig.~\ref{fig:Medusa}a at 7 and 8~K. The slope of the fits is
proportional to $B$, which we observed did not strongly depend on
the temperature for any of these samples. The intercept
$\gamma_{th}$ was observed to increase with $T$ due to the greater
thermal quasiparticle population. We shall examine each quantity
in turn.

In order to determine the value of $B$, an estimate of $n_{ph}$
for a given $\Phi$ is required, which we will conveniently pick as
the saturation fluence. The average energy density deposited by
the pump beam within the penetration depth at saturation is
$\Phi_{sat}/\lambda = 9.6~{\rm J/cm^3}$, using the value for
$\Phi_{sat}$ determined above. We compare this value with an
estimate based on the assumption that saturation occurs when the
quasiparticle system receives an energy comparable to the
condensation energy, which we approximate from the BCS theory as
$1/2 N(E_f)\Delta^2$. Here, $N(E_f)$ is the density of states at
the Fermi level and $\Delta$ is the superconducting gap energy.
Using an average gap size of $\Delta$=10~meV and $N(E_f)=7.3~{\rm
eV^{-1}}$ per unit cell \cite{ding}, we obtain a condensation
energy of ${\rm 0.3~J/cm^3}$. This value is significantly smaller
than the experimentally measured saturation energy of $9.6~{\rm
J/cm^3}$. Hence, only $3\%$ of the energy goes into creating
quasiparticle pairs. Using this value to convert the laser fluence
into excitation density, we obtain $B={\rm 2.48 \times
10^{-9}~cm^{3}/s}$. For a BCS superconductor in the dirty limit,
$B$ is proportional to the ratio of the electron phonon coupling
constant to the density of states at the Fermi energy
\cite{kaplan}. In the case of pnictide superconductors with
multiple gaps, a detailed theoretical model for the quasiparticle
recombination is necessary to further interpret the measured value
of $B$.

Figure~\ref{fig:Medusa}b presents $\gamma_{th}$ as a function of
temperature for $x=0.4-0.6$. Significantly, the data presented in
this Figure reveal that the thermal decay rate $\gamma_{th}=B
n_{th}$ is proportional to $T^2$ below $T_c$ for these doping
levels. This suggests the presence of a node in the gap around
which $n_{th} \sim T^2$, as opposed to the exponential dependence
expected from an isotropic gap. However, the linear dependence of
the magnitude of $\Delta R/R$ on $\Phi$ at weak excitation levels
(Fig.~\ref{fig:saturation}a) argues against the presence of nodal
quasiparticles, which would lead to $\Delta R/R \propto
\Phi^{2/3}$, as discussed above.

\begin{figure}
\includegraphics[scale=1]{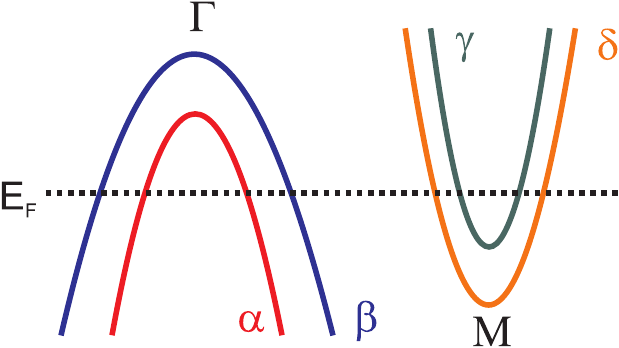}
\caption{\label{fig:FS} Schematic representation of the pnictide
band structure. While five bands have been observed by ARPES, the
inner hole band is nearly doubly degenerate and so is represented
by the single band $\alpha$.}
\end{figure}

The phenomenological consistency of the pump-probe measurements in
the doping range $x=0.4 - 0.6$ indicate that the physics of these
materials is consistent with the band-dependent picture we have
developed for the $x=0.4$ material\cite{torchinsky}. As this
picture is central to our understanding of nonequilibrium
quasiparticle dynamics in the pnictides, we briefly summarize it
using the simplified representation of a pnictide band structure
in Fig.~\ref{fig:FS}.

LDA calculations\cite{ma} have shown a lack of states 1.5 eV above
and (accounting for renormalization\cite{wray}) below the Fermi
level at the M point of the Brillouin zone, implying that the
probe wavelength used in these experiments is not able to couple
to quasiparticle dynamics at this location in the zone. This point
is bolstered by an estimate of the Sommerfeld parameter, which
shows the density of states of the hole bands to be several times
larger than at the electron pockets\cite{ding}, resulting in a
much greater number of photoinduced quasiparticles at the hole
pockets as compared with the electron pockets.

Thus, the electron bands $\gamma,\delta$ do not contribute
significantly to signal, i.e., $\Delta R(t)/R$ derives almost
exclusively from the hole bands. In particular, strong coupling
between the carriers in the $\alpha$ bands with those in
$\gamma,\delta$\cite{richard,wray} leads to the observed
excitation density dependent dynamics, while the isolation of
carriers in $\beta$ from the other bands is represented by their
slow, bottlenecked decay. Furthermore, since the entirety of the
signal is due to the hole bands, we may thus interpret the
observation $\Delta R(0)/R\propto\Phi$ to indicate that it is
these bands which possess fully gapped excitations.

As the excitation density is lowered, the thermal quasiparticle
population becomes comparable to, and then exceeds, the
photoinduced one. Nonequilibrium quasiparticle recombination is
then dominated by the second term in Eq.~\ref{eq:rt-lin}, which
represents a photoinduced quasiparticle recombining with a thermal
one. This measurement thus becomes particularly sensitive to the
presence of either a highly anisotropic gap or a node on the Fermi
surface, as the locally small gap energy leads to a relatively
higher concentration of thermal quasiparticles in its vicinity.

Here, the observation that $\gamma_{th}\propto T^2$ may be
indicative of either a highly anisotropic gap or node on the
electron pockets, as interband scattering may be dominated by
photoexcited carriers in $\alpha$ recombining with a relatively
large population of thermally present carriers in $\gamma,\delta$.
This scenario is compatible with an $s_{\pm}$ order parameter with
impurity scattering to account for the $T^2$ behavior, although it
does not require it. The fact that $\gamma_{th}\propto T^2$ does
not change up to doping levels of $x=0.6$ suggests that the Fermi
surface topology is consistent up to this doping.

\subsection{\label{sec:bakfeasu}Ba$_{1-x}$K$_x$Fe$_2$As$_2$: The High Fluence Regime}

\begin{figure}
\includegraphics[scale=1]{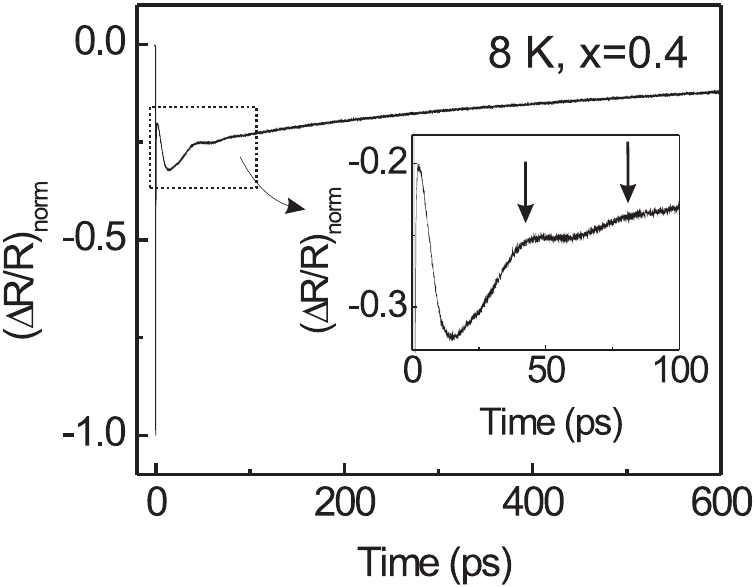}
\caption[Raw data]{\label{fig:brillouin} Dynamics of the decay
measured for a longer time window at $T=8$~K and $\Phi=37~{\rm \mu
J/cm^2}$. Highly damped oscillations are observed with a period of
$\sim 40$~ps (inset).}
\end{figure}

We now shift our attention to phenomena that require a different
framework than that developed above, i.e., the observation of
coherent oscillations in all samples at the highest fluences
($\Phi > 10 \mu$J/cm$^2$) and the dynamics of the underdoped
hole-doped samples where $x=0.2,0.3$.

At fluences greater than $10 \mu$J/cm$^2$, we observe the onset of
highly damped oscillations in the short time transients shown in
Figs~\ref{fig:rawdata}b and \ref{fig:rawdata}c that become more
apparent with increasing fluence. A clearer depiction of this
effect is provided by Fig.~\ref{fig:brillouin} by extending the
measurement window for the $x=0.4$ sample. These oscillations are
due to stimulated Brillouin Scattering \cite{thomsen}, where
interference between the portion of the probe beam reflected from
the sample surface and from the propagating strain pulse launched
by the pump modulates the signal in the time domain. The observed
period of $\sim$ 40~ps is consistent with the expected value of
$\lambda_{pr}/(2nv_s \cos\theta)$ \cite{thomsen} where
$\lambda_{pr}$ is the wavelength of the probe, $\theta$ the angle
of incidence, $v_s$ the speed of sound \cite{zbiri} and $n$ the
refractive index. The fast damping rate $\Gamma$ is set by the
small optical penetration depth of the light
($\Gamma=v_s/\lambda$).

The strength of the SBS signal is linearly proportional to the
thermal expansion coefficient. Thus, the disappearance of the
oscillations in the low fluence data below $T_c$ is a
manifestation of its large drop at the transition temperature
$T_c$ for all Ba-122 pnictides. This drop is due to the
thermodynamic relationship between $c_p(T)$ and the
expansivity\cite{budko,hardy,daluz} and is more pronounced for
expansion along the $c$-axis as a result of the layered nature of
the pnictides\cite{daluz}. The presence of an acoustic wave may be
an indication that the sample has been partially driven into the
normal state, even at the relatively modest fluences used here
(${\rm\leq 37~\mu J/cm^2}$).

We examine the plausibility of this scenario by simple energy
conservation arguments. We recall that the experimentally measured
saturation energy, estimated above as $9.6~{\rm J/cm^3}$, supports
the hypothesis that high fluences deplete the superconducting
condensate. However, this also indicates that a large fraction of
the energy deposited by the laser is transferred into other
internal degrees of freedom or is carried out of the probed region
by through-plane transport of the excited electrons. The large
excitation gradient that results from the short penetration depth
of the pump may result in fast diffusion of the excited electrons,
quickly transporting a significant fraction of the absorbed energy
out of the experimentally probed region.

\begin{figure}
\centering
\includegraphics[scale=1]{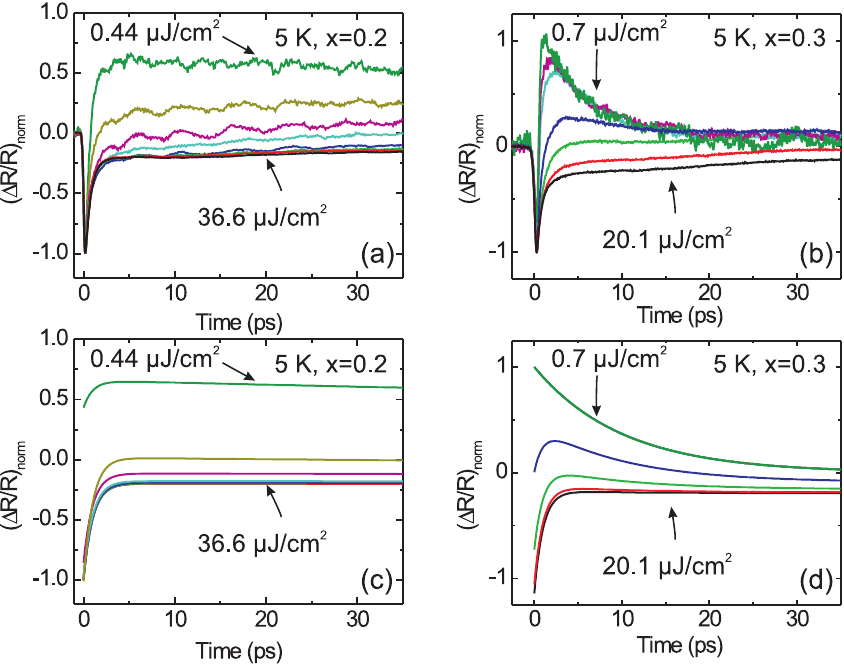}
\caption{\label{fig:under_hole_doped_raw}Raw traces in the
underdoped hole-doped samples Ba$_{1-x}$K$_x$Fe$_2$As$_2$ for (a)
$x=0.2$ ($T_c = 22$~K, $T_{SDW} \sim 100$~K) and (b) $x=0.3$ ($T_c
= 22$~K) at $T = 5$~K. The absorbed fluences are $\Phi$ = 36.6
(black -- (a) only), 20.1, 12.7, 7.0, 4.4, 2.2, 1.4, and 0.7
(blue) ${\rm \mu J/cm^2}$. Also shown are simulated results for
$\Delta R(t)/R$ at various values of $\Phi$ using a simple
two-component model in the (c) $x=0.2$ sample and the (d) $x=0.3$
sample at the same fluences. The simulations, which are performed
without free parameters, show excellent qualitative agreement with
the data.}
\end{figure}

Shifting focus to the underdoped samples,
Figs.~\ref{fig:under_hole_doped_raw}a and
\ref{fig:under_hole_doped_raw}b show the $T=5$~K reflectivity
transients in the $x=0.2$ and $x=0.3$ samples. In the $x=0.2$
material, $T_c = 22$~K and a spin-density wave transition occurs
at $T_{SDW} = 120$~K. In the $x=0.3$ sample, $T_{SDW}$ is
suppressed to 100~K and $T_c$ is elevated to $32$~K\cite{chen2}.
The dynamics observed here are markedly different than those of
the optimally and overdoped samples. In the $x=0.2$ compound, the
high fluence traces resemble the normal state of the $x = 0.4-0.6$
samples. With reduced fluence, we see the emergence of an
intensity-independent, slowly-decaying positive component
reminiscent of the slow dynamics of Fig.~\ref{fig:slow}a.

The dynamics of the $x=0.3$ sample shown in
Fig.~\ref{fig:under_hole_doped_raw} reveal behavior similar to
those of the $x=0.2$ sample at the highest fluences, again in
contrast with the measurements of Figs.~\ref{fig:rawdata}a-c.
Here, we observe that with a reduction of the incident laser
fluence, a positive component of the signal emerges whose recovery
dynamics are intensity independent and decay on the relatively
quick $\sim 10$~ps timescale.

\begin{figure}
\includegraphics[scale=1]{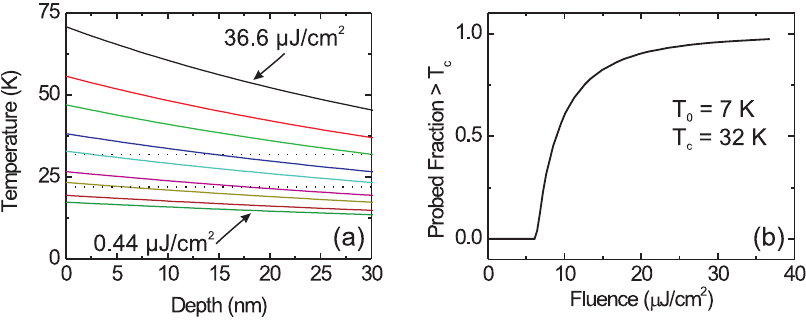}
\caption{\label{fig:sat_model} (a) Simulated temperature profile
as a function of depth for the fluences used in this study. The
dotted lines indicate the values of transition temperature for the
$x=0.2$ ($T_c=$ 22~K) and the $x=0.3$ ($T_c=32$~K) samples. (b)
Fraction of the sample heated into the normal state as
interrogated by the penetration depth of the probe for the $x=0.3$
sample.}
\end{figure}

The absence of a bimolecular intensity dependence may be
interpreted within the framework of the Rothwarf-Taylor equations
as bottlenecked recombination. Unfortunately, this precludes an
analysis to determine gap symmetry. However, we may understand
these data by applying a simple model which considers the
quasi-equilibrium temperature of the sample $T(z)$ as a function
of depth in the sample $z$ after excitation, the results of which
are shown in Figs.~\ref{fig:under_hole_doped_raw}c and
\ref{fig:under_hole_doped_raw}d. Conservation of energy provides
an expression for $T(z)$:
\begin{equation}\label{eq:temp}
\frac{\Phi}{d}\exp{\left(-z/d\right)}=\int_{T_0}^{T(z)}c_p(T)dT.
\end{equation}
Here, the specific heat $c_p(T)$ was modelled as a fifth-order
polynomial using published data on the $x=0.4$ compound\cite{ni}.
Equation~\ref{eq:temp} was then integrated and inverted
numerically to reproduce the initial temperature profile, depicted
in Fig.~\ref{fig:sat_model}a. We note that high fluences produce a
significant increase in temperature. In contrast, this increase is
very modest at the low fluences examined in
Sec.~\ref{sec:bakfeaso}.

The pump-probe measurement is an integrated measurement in depth.
We therefore calculate the fraction $A$ of the sample which has
been brought above the transition temperature from
\begin{equation}\label{eq:scfrac}
A=\frac{\int_0^{\infty}\Theta(z')\exp{\left(-2z'/d\right)}dz'}{\int_0^{\infty}\exp{\left(-2z'/d\right)}dz'},
\end{equation}
where $\Theta(z') = 1$ if $T(z')\geq T_c$ and $=0$ otherwise. The
factor of two in the above equation accounts for the reflection of
the beam back through the sample. In practice, this integral was
numerically evaluated to 3 penetration depths. The result is shown
in Fig.~\ref{fig:sat_model}b. For an initial temperature of 7~K,
the entire sample remains below $T_c$ for ${\rm \Phi<6~\mu
J/cm^2}$ and achieves a 90:10 distribution between normal and
superconducting state at ${\rm \Phi<19.8~\mu J/cm^2}$. The small
discrepancy between the value obtained from this simple model and
the experimentally measured saturation fluence may be due to
through-plane electronic diffusion mentioned above.

We consider the signal to comprise two parts. The first is a fast
decaying negative component of timescale $\tau_f$ with an offset
$c$ as $f(t)=-(\exp{\left(-t/\tau_f\right)}+c)$ to represent the
normal state contribution. The second is a slow, positive
component for the superconducting contribution of timescale
$\tau_s$ as $s(t)=\exp{\left(-t/\tau_s\right)}$. The two
components are then added in the proportion $Af(t) + (1-A)s(t)$,
where the coefficient $A$ is defined as the fraction of the sample
seen by the probe beam with $T(z)>T_c$. Here, we compute $A$ as a
weighted average of the temperature, as in Eq.~\ref{eq:scfrac}.

The results of this calculation using $c=0.2$, $\tau_f=1$~ps and
$\tau_s=400$~ps are shown in Fig.~\ref{fig:under_hole_doped_raw}c
for the $x=0.2$ sample. Fig.~\ref{fig:under_hole_doped_raw}d uses
the same vales for $c$ and $\tau_f$ but a value of $\tau_s=10$~ps
for the $x=0.3$. In both cases, we observe excellent qualitative
agreement between the experimental data and the model using no
fitted parameters. Data acquired at higher initial temperatures
showed similar agreement with this model for both $x=0.2$ and
$x=0.3$.

We note that this model is not able to replicate the dynamics of
the optimally doped and overdoped samples discussed in the
previous section. Even when accounting for the change in sign of
the phenomenological superconducting component $s(t)$ from
positive to negative, no combination of simulation parameters can
produce an initial decay rate that depends linearly on excitation
fluence. This provides further evidence that the intensity
dependence observed in the $x=0.4-0.6$ samples arises from
bimolecular recombination kinetics. We thus emphasize that this
analysis has no impact on the analysis performed in the
low-fluence regime. First, in those traces, we deduced the gap
symmetry from the initial quasiparticle decay rate $\gamma_0$,
which was obtained from the slope of the reflectivity transients
immediately after the peak, i.e., $t<1$~ps. This is much shorter
than the time required for an equilibrium temperature to be
established between the electronic system and the phonons. Second,
all the data used in the analysis of Sec.~\ref{sec:bakfeaso} were
at such low fluences that the simulated equilibrium sample
temperature did not depart significantly from the initial value.

The physical processes at play in the $x=0.2$ and $x=0.3$ samples
may also be understood within the framework of the band-dependent
dynamics discussed in the previous section. Ultrafast measurements
on the Ba$_{1-x}$K$_{x}$Fe$_2$As$_2$ samples for values of $x<0.3$
indicate a competition for itinerant electrons between the SDW
state and the superconducting state\cite{chia}. This observation
is consistent with ARPES measurements on
Ba$_{0.75}$K$_{0.25}$Fe$_2$As$_2$ which have shown that the
superconducting fraction in $\alpha$ and $\gamma,\delta$ are
suppressed due to their contribution to SDW order via Fermi
surface nesting\cite{xu}. At $x=0.2$, this nesting is even
slightly more favored and hence the bulk of the optical pump-probe
response in the superconducting state arises from the slow,
intensity independent decay in the outer hole band $\beta$ with
little to no contribution to the signal from the other bands.

In the $x=0.3$ sample, the absence of intensity dependence is
again due to the presence of a SDW state between $\alpha$ and
$\gamma,\delta$ which competes with superconductivity for carriers
in these bands. It is, however, unclear at present why the
dynamics change from slow to fast at this particular doping.

\begin{figure}
\centering
\includegraphics[scale=1]{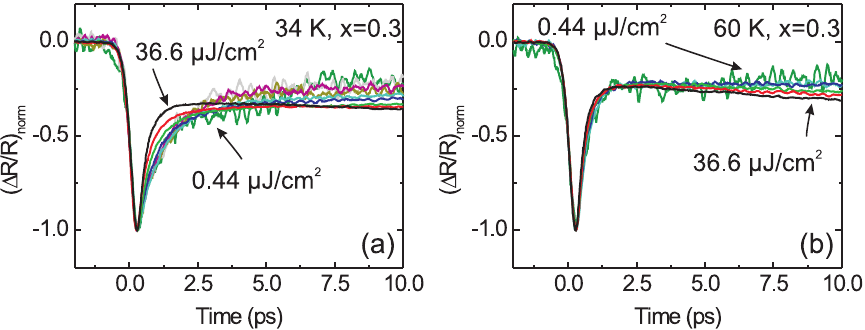}
\caption{\label{fig:pseudogap_raw}Raw traces in the underdoped
hole-doped pnictide Ba$_{0.7}$K$_{0.3}$Fe$_2$As$_2$ ($T_c = 32$~K)
above $T_c$. In (a), we note that there is some residual intensity
dependence of the signal, where the initial recovery of the fast
component is quicker at the highest fluences than at lower ones.
This is identical to the superconducting state, despite the fact
that at $T=34$~K, the sample is in the normal state and the SDW
state does not exhibit intensity dependent recombination (see
below). After the initial electronic recovery, the amplitude of
the acoustic wave is diminished at lower fluences. (b) In
contrast, the fluence dependence of the electronic recovery has
all but disappeared at 60~K. In both plots, the absorbed fluences
are $\Phi$ = 36.6, 20.1, 12.7, 7.0, 4.4, 2.2, 1.4, 0.7, and 0.44
${\rm \mu J/cm^2}$. }
\end{figure}

The presence of a pseudogap state in the 1111 system has already
been argued for based on ultrafast pump-probe
measurements\cite{mertelj}; ARPES\cite{xu} and optical
conductivity\cite{kwon2010} have indicated the same for underdoped
Ba$_{1-x}$K$_x$Fe$_2$As$_2$ samples. In the context of the
intensity dependent dynamics studied here, pseudogap behavior is
suggested by the intensity dependence present above $T_c$ in
Fig.~3c of Ref.~\onlinecite{torchinsky} for $x=0.4$. A clearer
demonstration of such fluence dependence is depicted in the
time-resolved reflectivity transients of
Fig.~\ref{fig:pseudogap_raw}a, which were measured in the $x=0.3$
sample 2~K above the transition temperature. There is a
discernable tendency towards faster initial relaxation of the
reflectivity transients with an increase in temperature. This
disappears by 60~K, as may be seen in
Fig.~\ref{fig:pseudogap_raw}b.

In light of the band-dependent relaxation dynamics discussed
above, we posit that those carriers participating in the pseudogap
state must derive from the $\alpha$ band because these carriers
give rise to intensity dependent recombination in the
superconducting state. This conclusion is consistent with prior
ARPES measurements on Ba$_{0.75}$K$_{0.25}$Fe$_2$As$_2$ which show
the presence of a pseudogap in $\alpha$ but not in the other
bands\cite{xu}.

\subsection{\label{bafecoas}BaFe$_{2-y}$Co$_y$As$_2$}

\begin{figure}
\includegraphics[scale=1]{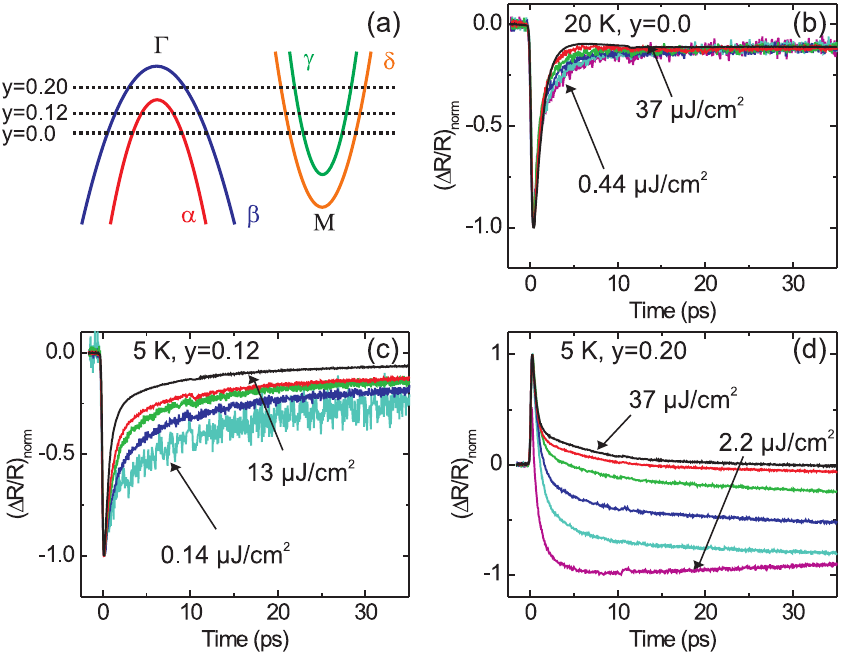}
\caption{\label{fig:elec_doped_raw} (a) Schematic representation
of the band structure of the Ba-122 pnictides. Although this is
somewhat of a simplification that ignores the change in $k_z$
dependence with doping\cite{thir}, the predominant effect is a
shift of the Fermi level up with the addition of
electrons\cite{brouet}: in the parent compound, $\alpha$ and
$\gamma, \delta$ nest, leading to SDW behavior. In the underdoped
regime ($y=0.12$), $\alpha$ does not nest appreciably with
$\gamma,\delta$. With the addition of more holes, the $\alpha$
band is submerged below the Fermi level when the doping reaches
$y=0.15$, as measured by Sekiba et al.\cite{sekiba} (b) Raw
$(\Delta R/R)_{norm}$ traces for the parent compound at 20~K. The
moderate amount of intensity dependence in these traces at short
times is due to acoustic suppression from the diminished thermal
expansion parameter at lower temperatures/fluences. (c)
Reflectivity transients in the underdoped sample ($y=0.12$,
$T_c=12$~K) show a mild amount of intensity dependence of the
faster relaxation with increasing fluence, similar to that seen in
the hole-doped samples. (d) Fluence dependence of the optimally
doped sample ($y=0.2$, $T_c = 22.5$~K) indicates entirely
different relaxation dynamics with fluence. While there appears to
be intensity dependence present here, the slope of the initial
transient, $\gamma_0$ is identical for all fluences. The
differences between the various fluences and the origin of this
non-uniformity of the relaxation dynamics are due to a temperature
effect, as described in the text.}
\end{figure}

We now shift our focus to the electron doped side of the phase
diagram. Fig.~\ref{fig:elec_doped_raw}a depicts the qualitative
effect of doping on the Fermi level. As electrons are added to the
system, the Fermi level rises in a rigid band manner, as has been
observed by Brouet et al.\cite{brouet} At a Co-doping of $y=0.15$,
the $\alpha$ band begins to submerge below the Fermi level at
various points along $k_z$\cite{sekiba}. By the point of optimal
Co-doping at $y=0.20$, $\alpha$ has been submerged well below the
Fermi level.

The reflectivity transients of Fig.~\ref{fig:elec_doped_raw}
exhibit the consequences of the disappearance of $\alpha$. As a
point of comparison, Fig.~\ref{fig:elec_doped_raw}b shows data
from the parent compound taken at 20~K, well below the SDW
transition temperature $T_{SDW}=143$~K. None of the intensity
dependent relaxation dynamics associated with the $x>0.4$
hole-doped samples were observed in the undoped sample. We note
the small effect of the suppression of the thermal expansion
predominantly along the $c$-axis\cite{hardy,daluz,budko} which
inhibits the generation of an acoustic wave. Qualitatively
identical behavior was observed in an underdoped sample ($y=0.08$,
not shown) for which the SDW transition is suppressed to 95~K but
which does not superconduct at low temperatures.

Fig.~\ref{fig:elec_doped_raw}c shows raw data for ${\rm
BaFe_{1.88}Co_{0.12}As_2}$ ($T_c=12$~K) taken at different
fluences ranging from 0.14~${\rm \mu J/cm^2}$ to 13~${\rm \mu
J/cm^2}$. As with the $x=0.4-0.6$ samples, we observe a fast
intensity dependence of the signal at short times with higher
laser intensities corresponding to faster initial recombination
rates. We note that the relatively low $T_c=12$~K of the $y=0.08$
sample does not allow us to use the initial decay-rate analysis
employed above to determine Fermi surface topology.

With the submergence of the inner hole band $\alpha$ at
$y=0.15$\cite{sekiba}, the characteristics of $(\Delta
R/R)_{norm}$ change markedly. Although qualitatively similar
behavior was observed in the $y=0.16$ ($T_c=16$~K) sample, these
dynamics are most clearly represented by the $y=0.20$ sample due
to its slightly higher $T_c= 22$~K. As may be seen in
Fig.~\ref{fig:elec_doped_raw}d, $\Delta R/R$ traces scaled to the
initial electronic response at $t=0$ are qualitatively identical
(within a sign) to those of the $x=0.2$ underdoped hole-doped
sample in Fig~\ref{fig:under_hole_doped_raw}a. Indeed, we observe
that the polarity of the signal has changed from the underdoped to
the optimally doped side of the phase diagram. A similar effect
has also been observed in the ${\rm
Bi_2Sr_2Ca_{1-y}Dy_yCu_2O_{8+d}}$ (BSCCO) system of cuprate
superconductors\cite{gedik3}.

\begin{figure} \includegraphics[scale=1] {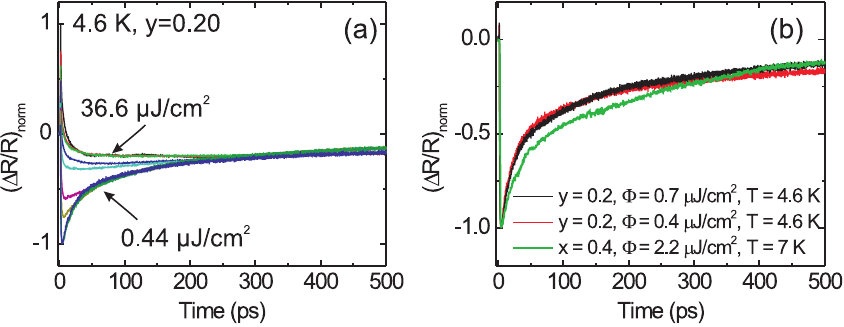}
\caption{\label{fig:compare} Evolution of $(\Delta R/R)_{norm}$ in
${\rm BaFe_{1.8}Co_{0.2}As_2}$ for absorbed fluences of $\Phi$ =
36.6 (black), 20.1, 12.7, 7.0, 4.4, 2.2, 1.4, 0.7, and 0.44 (blue)
${\rm \mu J/cm^2}$. As the fluence is increased, the recovery to
equilibrium seems to slow due to a greater fraction of the probed
sample being heated above $T_c$, especially at these long time
scales. Proper scaling of the various traces yields a common,
bottlenecked decay, reached by the sample when the entire probed
region has cooled below $T_c$ due to thermal diffusion. (b)
Comparison of long-time traces of the optimally-doped hole-doped
sample (${\rm Ba_{0.6}K_{0.4}Fe_2As_2}$) and the optimally
electron-doped (${\rm BaFe_{1.8}Co_{0.2}As_2}$) material at
fluences of 0.7 and 0.4 ${\rm \mu J/cm^2}$. Even though the Fermi
surface topologies and gap energies of the two samples differ
markedly, we note that the relaxation dynamics are nearly
identical, indicating a common underlying relaxation mechanism,
i.e., bottlenecked quasiparticle recombination in the outer hole
band ($\beta$). This slow relaxation switches off at the
respective values of $T_c$ of both samples.}
\end{figure}

In order to better interpret the dynamics of
Fig.~\ref{fig:elec_doped_raw}d, we modelled the heat flow out of
the excited region after the establishment of an initial
equilibrium temperature. This was done by numerically solving the
heat diffusion equation using literature temperature-dependent
thermal conductivity\cite{Checkelsky} and heat capacity\cite{ni}
data while assuming a temperature-independent density. The initial
condition was specified using Eq.~\ref{eq:temp} and the incident
fluence. In the limit of bottlenecked recombination, $\Delta
R(t)/R$ should be independent of initial quasiparticle density
once the entire probed region is below $T_c$. This indicates that
the data should be scaled at the time $t$ when the entire sample
becomes superconducting.

Reflectivity transients were acquired out to 600~ps and then
rescaled at the point in time at which the entire probed depth is
below $T_c$. This rescaling is shown in Fig.~\ref{fig:compare}a,
where we observe a remarkable convergence of the data at long
times. The highest fluences measured produce the normal state
response, i.e., an initial ``spike" followed by a quasi-steady
state that does not decay within the time window. As the
excitation fluence is reduced, the initial recovery gradually
transitions to bottlenecked quasiparticle recombination at earlier
and earlier times. At the lowest fluences (${\rm \Phi = 0.7~\mu
J/cm^2}$ and ${\rm \Phi = 0.44~\mu J/cm^2}$), the entire probed
region is below $T_c$ and there is an intensity independent,
bottlenecked relaxation of the signal.

This behavior recalls the slow dynamics of the optimally
hole-doped sample shown in Fig.~\ref{fig:slow}a. A comparison of
these two data sets is provided in Fig.~\ref{fig:compare}b which
plots data from the optimally hole-doped sample ($x=0.4$) at
$\Phi=$~2.2~${\rm \mu J/cm^2}$ alongside data recorded at the
lowest fluences in the optimally electron-doped sample ($y=0.2$).
We observe nearly identical behavior, suggesting that these decays
arise from the same set of dynamics. As mentioned above, this slow
decay was observed to switch off in both samples abruptly at their
respective superconducting transition temperatures, indicating
that it originates from superconductivity.

We posit that the similarities between the long-time reflectivity
transients produced by Ba$_{0.6}$K$_{0.4}$Fe$_2$As$_2$ and
BaFe$_{1.8}$Co$_{0.2}$As$_2$ may also account for the differences
between the data in Figs.~\ref{fig:elec_doped_raw}c and
\ref{fig:elec_doped_raw}d. That is, the disappearance of
bimolecular recombination with the doping of electrons derives
from the band-dependent recombination dynamics presented above in
Sec.~\ref{sec:bakfeaso}. The addition of electrons disrupts the
near-perfect nesting of $\alpha$ with $\gamma,\delta$, weakening
the effect of the magnetic resonance on Cooper pairing and
resulting in a diminished intensity dependence of the $y=0.12$
sample of Fig.~\ref{fig:elec_doped_raw}c. Further doping of
electrons leads to the submergence of $\alpha$ and thus the
remaining relaxation in the $y=0.20$ sample is the slow decay of
photoinduced carriers in $\beta$.

\section{\label{end}Summary and Conclusion}

We have presented an ultrafast pump-probe study of the Ba-122
pnictides as a function of fluence and doping on both sides of the
phase diagram that provides evidence for band-dependent dynamics.
For optimally doped Ba$_{1-x}$K$_x$Fe$_2$As$_2$, the initial
recovery rate of the reflectivity transients depends linearly on
the incident laser fluence $\Phi$ in the low excitation limit.
This component of the signal was argued to derive from the inner
hole band $\alpha$. A Rothwarf-Taylor analysis of this intensity
dependence in the low-fluence limit yields a $T^2$ behavior for
the thermal population of quasiparticles. At long times, $\Delta
R/R$ tended towards a slow, intensity independent decay which
originates from the outer hole band, $\beta$. These observations
were unchanged with an increase of K-doping up to levels of
$x=0.6$. Our results thus indicate fully gapped hole bands with
nodal or strongly anisotropic electron bands in these samples.

In underdoped samples ($x=0.2, 0.3$), the intensity dependence
disappears as electrons in $\alpha$ are taken up by SDW ordering
and hence unable to participate in superconductivity. The fluence
dependence of $\Delta R(t)/R$ is then governed by the decay in
$\beta$ in a proportion which may be estimated by considering the
fraction of the probed depth which is driven above $T_c$ by the
pump.

In the BaFe$_{2-y}$Co$_y$As$_2$ samples, we observed intensity
dependent recombination at low doping levels ($y=0.12$) which
completely disappeared with the submergence of the inner hole band
$\alpha$ by the doping of electrons. The long-time dynamics of
these samples were consistent with bottlenecked behavior due to
quasiparticle recombination in the outer hole band $\beta$.

Intensity dependence was observed to persist above the transition
temperature $T_c$ for samples with K-doping at levels of $x\leq
0.4$ and was taken as evidence for a pseudogap state. This
pseudogap behavior was posited to arise from $\alpha$, consistent
with prior ARPES measurements. No such behavior is observed on the
electron doped side of the phase diagram.

Stimulated Brillouin scattering was observed in all samples above
the superconducting transition temperature but disappeared
immediately below the transition temperature due to a large
decrease of the thermal expansion coefficient at $T_c$. The
presence of Brillouin Scattering below the transition temperature
for the highest incident fluences used in this study (e.g., $\sim
{\rm 37~\mu J/cm^2}$) is an indication that, even at these
relatively modest fluences, the optical pulse was able to bring
the sample into the normal state within the penetration depth of
the excitation. At the highest fluences, this picture correlated
well with basic modelling of reflectivity transients in terms of a
linear combination of a normal-state response and a
superconducting, bottlenecked quasiparticle recombination, the
relative proportions of which are dictated by a simple equilibrium
temperature model. In the low fluence regime, however, where no
part of the probed region could reach a temperature greater than
$T_c$, either only bottlenecked or bare bimolecular recombination
were observed.

\section{Acknowledgments}

The authors thank Prof. B. Andrei Bernevig and Alex Frenzel for
useful discussions. This work was supported by DOE Grant No.
DE-FG02-08ER46521, the MRSEC Program of the National Science
Foundation under award number DMR - 0819762, the NSFC, CAS and the
973 project of the MOST of China.

\bibliography{feas}

\end{document}